# A Hierarchical Context-aware Modeling Approach for Multi-aspect and Multi-granular Pronunciation Assessment

*Fu-An Chao[1], Tien-Hong Lo[1, 2], Tzu-I Wu[1, 2], Yao-Ting Sung[3], Berlin Chen[2]*

[1]Research Center for Psychological and Educational Testing, National Taiwan Normal University
[2]Department of Computer Science and Information Engineering, National Taiwan Normal University
[3]Department of Educational Psychology and Counseling, National Taiwan Normal University
`{fuann, teinhonglo, zoetwu, sungtc, berlin}@ntnu.edu.tw`

## Abstract

Automatic Pronunciation Assessment (APA) plays a vital role in Computer-assisted Pronunciation Training (CAPT) when evaluating a second language (L2) learner's speaking proficiency. However, an apparent downside of most de facto methods is that they parallelize the modeling process throughout different speech granularities without accounting for the hierarchical and local contextual relationships among them. In light of this, a novel hierarchical approach is proposed in this paper for multi-aspect and multi-granular APA. Specifically, we first introduce the notion of sup-phonemes to explore more subtle semantic traits of L2 speakers. Second, a depth-wise separable convolution layer is exploited to better encapsulate the local context cues at the sub-word level. Finally, we use a score-restraint attention pooling mechanism to predict the sentence-level scores and optimize the component models with a multitask learning (MTL) framework. Extensive experiments carried out on a publicly-available benchmark dataset, viz. speechocean762, demonstrate the efficacy of our approach in relation to some cutting-edge baselines.

**Index Terms**: Automatic pronunciation assessment, computer-assisted pronunciation training, multi-task learning

## 1. Introduction

In response to the surging demand for self-directed language learning, Computer-assisted Pronunciation Training (CAPT) has emerged as a prevailing alternative for second language (L2) learners. In addition, CAPT systems also figure prominently in supplementing the teacher's instruction, meanwhile alleviating their workload. Typically, a CAPT system is deployed in a "read aloud" scenario where the text prompt is presented and an L2 learner is asked to pronounce it aloud and learn the manner of speech of native speakers. Through persistent repetition and practice, it is anticipated that non-native speakers can steadily improve their pronunciation proficiency. According to the types of feedback, CAPT can be roughly divided into two categories: one is mispronunciation detection and diagnosis (MDD) and the other is automatic pronunciation assessment (APA).

MDD aims to precisely detect the erroneous articulations produced by L2 learners and provide the correct diagnosis. The detection target usually focuses exclusively at the segmental-level (e.g., phone). Among traditional pipeline approaches, most of the endeavors on MDD typically leverage an automatic speech recognition (ASR) system to obtain phone-level pronunciation quality, so as to pinpoint L2 learners' mispronunciations. Such decisive quality measures like Goodness of Pronunciation (GOP) [1] and its variants [2][3] are the most prevalent instantiations. In general, they compute a ratio between the likelihoods of a canonical phone sequence and the most likely pronounced phones given the learner's actual speech sound. Subsequently, phone-level mispronunciations are detected by given phone-dependent thresholds. Recently, there has been a growing interest in the adoption of end-to-end (E2E) based methods [4][5][6], which view MDD as a free-phone recognition process. By comparing the dictation result with the canonical phone sequence of the prompt, erroneous pronunciations can be readily identified. Among these methods, the wav2vec 2.0-based encoder [7] has achieved promising results and shown its superiority over previous attempts [8].

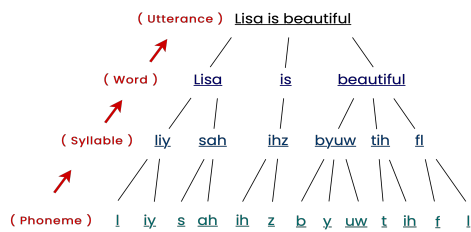

Figure 1: *A schematic diagram of the acoustic hierarchical structure*

Different from MDD, APA concentrates more on the scoring of non-native speakers' grasp of the language. Apart from the segmental features, suprasegmental features (e.g., fluency) are also considered crucial; this, however, makes APA even more challenging. Since it is hard to achieve high-level inter-rater agreements on these assessments, research on APA is much scantier than that on MDD. Although some efforts have been made on predicting fluency, lexical stress and intonation, the predictive scores of these aspects are usually modeled independently [9][10][11]. Other research directions explore different granular assessments; they however are usually centralized on either the word- or sentence-level assessment, and only a holistic score is presented [12][13]. Recently, in order to provide more comprehensive and fine-grained feedback, there is a trend to advocate the notion of multi-aspect and multi-granularity processing [14][15], which uses a unified model to perform different aspects and granular assessments simultaneously. Furthermore, these methods all follow a similar vein to exploit a BERT-based modeling paradigm [16] for learning utterance-level representations. Specifically, a set of trainable `[CLS]` tokens are introduced to predict the corresponding utterance-level scores. These tokens are spliced together with the phone-level input sequence and sent to the Transformer encoder. By leveraging the parallel attention mechanism, the model learns contextualized representations that correspond to each [CLS] token. However, while these individual representations are assigned to predict different utterance-level scores, they are embedded directly with the phone-level information. As shown in Figure 1, the speech signals are typically characterized by their intricate hierarchical composition where the relevant information at different granularities may be distinct. Therefore, when modeling different speech granularities in a BERT-alike fashion, it may be inconsistent with the intrinsic acoustic structure.

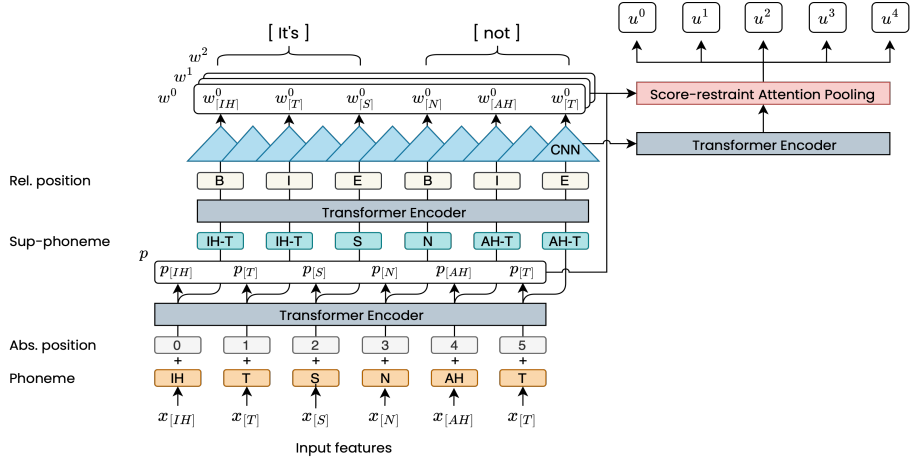

Figure 2. *A schematic diagram of proposed hierarchical modeling for automatic pronunciation assessment.*
$\{p\}$ refers to the phone-level accuracy scores,
$\{w^0, w^1, w^2\}$ refers to the word-level accuracy, stress, and total scores, respectively.
$\{u^0, u^1, u^2, u^3, u^4\}$ refers to the utterance-level accuracy, fluency, completeness, prosody, and total scores, respectively

On the other hand, the erroneous or unnatural pronunciations of non-native speakers are usually attributed to the articulation patterns of their mother tongues. In this regard, understanding the pronunciation patterns made by L2 learners or the cross-lingual articulated differences becomes pivotal to promoting the pragmatic utility of a CAPT system. For example, a novel anti-phone modeling [5] is proposed to generalize the categorical and non-categorical pronunciation errors occurring in L2 learners for better MDD performance. Due to the rhythmic difference between Chinese and English, [17] explores some rhythm measures based on sentence stress for L2 English rhythm evaluation. Furthermore, it is also indicated that the main source of the pronunciation errors in non-native Mandarin Chinese exist at sub-word level [18]. However, previous literature often omits the modeling at this level as well as the local context dependency between phonemes, which may inevitably cause the performance of the word-level assessments to be far from that of the human experts.

To tackle these problems, we in this paper propose a hierarchical context-aware modeling structure for APA. Through a comprehensive set of experiments conducted on a benchmark dataset, speechocean762 [19], we confirm that the proposed method can outperform previous methods on several assessment granularities. Strikingly, the experimental results also show that our method can yield significant improvements on word-level scoring.

## 2. Proposed Methodology

In this section, we will shed light on the proposed method which is disassembled into three parts: input acoustic features, hierarchical modeling, and optimization. An overview of the complete workflow is depicted in Figure 2.

### 2.1. Input acoustic features

Traditional GOP-based features can only offer the segmental-level information, which, however, is inadequate for portraying suprasegmental aspects. To relieve this constraint, we explore multi-view representations as those described in [15], which create a synergy of heterogeneous prosodic and self-supervised learning (SSL) based features to form the acoustic representations for the purpose of supporting cross-level assessments. The assembled acoustic representations are formulated as follows:

$$E_{multi} = [E_{gop}, E_{dur}, E_{eng}, E_{w2v2}, E_{hubert}, E_{wavlm}] \quad (1)$$

$$x_{acoustic} = Dense(E_{multi}) \quad (2)$$

where $E_{gop}$ denotes the conventional GOP; $E_{dur}$ and $E_{eng}$ refer to prosodic features of duration and energy, respectively; $E_{w2v2}, E_{hubert}$ and $E_{wavlm}$ are SSL-based features derived from wav2vec 2.0 [7], HuBERT [20], and WavLM [21]. Finally, these diverse features are concatenated and projected through a single-layer feed-forward dense network to obtain the acoustic representations $x_{acoustic}$ for the subsequent hierarchical modeling.

### 2.2. Hierarchical modeling approach

#### 2.2.1. Phone-level modeling

Apart from the acoustic representations, a common practice in building conventional CAPT systems is to inject phonological information embodied by a sequence of canonical phoneme embeddings of the text prompt [14][15]. Since different phonemes have different characteristics, the usage of canonical phone embedding can provide useful information when performing pronunciation assessments. Consequently, in phone-level modeling, we add the projected acoustic representations, canonical phone embedding, and an absolute positional embedding together and feed them into the Transformer encoder. The resulting phone-level representations are then used as input to a multi-layer perceptron (MLP) head to predict the phone-level scores $p$ (accuracy):

$$x_{phn} = x_{acoustic} + E_{phone} + E_{abs-pos} \quad (3)$$

$$h_{phn} = Transformer_{phn}(x_{phn}) \quad (4)$$

$$p = MLP(h_{phn}) \quad (5)$$

*2.2.2. Sub-word-level modeling*

Due to the limited phone set, the canonical phone embeddings lack the ability to capture contextual semantic information. To mitigate this issue, we capitalize on sup-phoneme embeddings to enhance the representation capacity and context awareness. Initially proposed in [22], sup-phone representations are created in conjunction with the canonical phone representations to enhance the quality of text-to-speech (TTS). Specifically, sup-phone is learned from the Byte-Pair Encoding (BPE) algorithm [23], which refers to a group of adjacent phones and does not necessarily correspond to a lexical word. Different from the original BPE algorithm, the learning process of sup-phone views each word as a sequence of phones rather than characters. Through the iterative merging process, the BPE model is trained until it reaches a pre-defined vocabulary size and can be used to encode a word into sup-phone level token(s). The whole encoding process is greedy and deterministic. To the best of our knowledge, this is the first attempt to use sup-phoneme in the field of CAPT. Further, sup-phone embeddings are jointly added with the phone level representations and modeled by the sub-word level Transformer encoder:

$$x_{sub-word} = h_{phn} + E_{sup-phone} \quad (6)$$

$$h_{sub-word} = Transformer_{sub-word}(x_{sub-word}) \quad (7)$$

*2.2.3. Word-level modeling*

As suggested in [18], Mandarin L2 speakers usually articulate wrong pronunciations at the sub-word level due to the unique tonal traits in the Chinese language. To capture such characteristics, we adopt a convolution neural network (CNN) to obtain a more contextualized word representation. First, we sum up each sub-word representation with a relative position embedding where the possible phone positions in a word are denoted by begin (B), internal (I), end (E), and single-phone word (S). Subsequently, this modified representation is modeled by a depth-wise separable convolution (DS-Conv) layer [24]. Since DS-Conv factorizes the standard convolution with the depth-wise convolution and the point-wise convolution, its most attractive merit is the low computational cost. The following two equations illustrate the modeling procedure:

$$x_{word} = h_{sub-word} + E_{rel-pos} \quad (8)$$

$$h_{word} = DSConv(x_{word}) \quad (9)$$

After the convolution operation, we can derive word-level representations and apply three MLP prediction heads to obtain the word-level scores $w^j$ (accuracy, stress and total):

$$w^j = MLP^j(h_{word}), \quad j = 0, 1, 2 \quad (10)$$

Note that, during the training stage, we propagate the word score to each of its phones. In the inference phase, the outputs of the MLP prediction heads that belongs to different words are averaged for consistency. In addition, since the word-level stress is a binary score (5 or 10), we use a sigmoid function to restrict the output value.

*2.2.4. Utterance-level modeling*

In utterance-level assessments, beyond using suprasegmental-level features, it is intuitive that the utterance-level scores should be correlated to the phone- and word-level scores. To make use of the intuition, we design a score-restraint attention-pooling mechanism and takes all the phone- and word-level scores to generate the attention weights for aggregating the hidden states of the utterance-level Transformer:

$$h_{utt} = Transformer_{utt}(h_{word}) \quad (11)$$

$$s_i = GELU(W([p_i, w_i^0, w_i^1, w_i^2]) + b), \\ i = 0, ..., N-1 \quad (12)$$

$$\alpha_i = \frac{e^{s_i}}{\sum_{i=0}^{N-1} e^{s_i}} \quad (13)$$

$$h_{agg} = \frac{1}{N} \sum_{i=0}^{N-1} \alpha_i * h_{utt}^i \quad (14)$$

The derived $h_{agg}$ is then fed into 5-layer MLP based prediction heads to obtain the corresponding utterance-level scores $u^k$ (accuracy, completeness, fluency, prosody, total):

$$u^k = MLP^k(h_{agg}), k = 0, 1, 2, 3, 4 \quad (15)$$

*2.2.5. Optimization*

The proposed hierarchical network is optimized based on the MTL framework where Mean Square Error (MSE) is utilized as the loss function and applied to each granularity. In this work, we average the losses of each granularity and simply sum them up as the ultimate loss function:

$$\mathcal{L}_{MTL} = \frac{1}{N_p} \sum_{i=0}^{N_p-1} \mathcal{L}_{p^i} + \frac{1}{N_w} \sum_{i=0}^{N_w-1} \mathcal{L}_{w^i} + \frac{1}{N_u} \sum_{i=0}^{N_u-1} \mathcal{L}_{u^i} \quad (16)$$

where $\mathcal{L}_{p^i}$, $\mathcal{L}_{w^i}$, and $\mathcal{L}_{u^i}$ are phone-level, word-level and utterance-level losses in $i_{th}$ aspect; $N_p$, $N_w$, and $N_u$ denote the numbers of aspect scores at phone, word, and utterance levels, respectively. In this work, $N_p = 1$, $N_w = 3$ and $N_u = 5$.

## 3. Experiments

### 3.1. Experimental setup

In this work, we utilized speechocean762 [19] to evaluate our proposed method. The dataset contains 5,000 English-speaking recordings by 250 Mandarin L2 speakers. One of its distinctive characteristics is rich sets of annotation information at different levels of granularity, including utterance-, word- and phone-level human gold-standard scores were equipped; all of them had been evaluated by five experts under the same rubrics. To facilitate training, all these scores are normalized into the same scale as the phone-level score which ranges from 0 to 2.

As for the model configuration, the multi-view representations $x_{acoustic}$ are extracted as the same configuration as [15]. The dimensions of $x_{acoustic}$ and $E_{abs-pos}$ are set to 24, $E_{phone}$, $E_{sup-phone}$, and $E_{rel-pos}$ are of 36 dimensions, and all the features and embeddings are projected to the same dimensionality of 24 for integration. In particular, the sup-phone embeddings are derived with the BPE algorithm which is learned from an open-source text corpus[1] with a base vocabulary size of 100. The layers of Transformer encoders at phone-, sub-word-, and utterance-level are set to 3, 1, 1, respectively. All of them have only 1 attention head and hidden vectors of 24 dimensions. In addition, DS-Conv has 72 kernels with the size of 3 for each, and the stride is set to 1. All the models were trained for 50 epochs with a batch size of 25. The initial learning rate was set to 1e-3 and halved every 5 epochs after the 20th epoch. To be consistent with the previous studies [14][15], we ran each experiment 5 times with different random seeds and reported the mean and standard deviation of Pearson Correlation Coefficient (PCC) and Mean Square Error (MSE).

Table 1. *Performance of different methods evaluated on speechocean762.*

| Model | Phone Score | | Word Score (PCC) ↑ | | | Utterance Score (PCC) ↑ | | | | |
|---|---|---|---|---|---|---|---|---|---|---|
| | MSE ↓ | PCC ↑ | Accuracy | Stress | Total | Accuracy | Completeness | Fluency | Prosody | Total |
| Lin et al. [13] | - | - | - | - | - | - | - | - | - | 0.720 |
| Kim et al. [25] | - | - | - | - | - | - | - | 0.780 | 0.770 | - |
| LSTM [14] | 0.089 ±0.000 | 0.591 ±0.003 | 0.514 ±0.003 | 0.294 ±0.012 | 0.531 ±0.004 | 0.720 ±0.002 | 0.076 ±0.086 | 0.745 ±0.002 | 0.747 ±0.005 | 0.741 ±0.002 |
| GOPT [14] | 0.085 ±0.001 | 0.612 ±0.003 | 0.533 ±0.004 | 0.291 ±0.030 | 0.549 ±0.002 | 0.714 ±0.004 | 0.155 ±0.039 | 0.753 ±0.008 | 0.760 ±0.006 | 0.742 ±0.005 |
| 3M [15] | 0.078 ±0.001 | 0.656 ±0.005 | 0.598 ±0.005 | 0.289 ±0.033 | 0.617 ±0.005 | 0.760 ±0.004 | 0.325 ±0.141 | 0.828 ±0.006 | 0.827 ±0.008 | 0.796 ±0.004 |
| HiPAMA [26] | 0.084 ±0.001 | 0.616 ±0.004 | 0.575 ±0.004 | 0.320 ±0.021 | 0.591 ±0.004 | 0.730 ±0.002 | 0.276 ±0.177 | 0.749 ±0.001 | 0.751 ±0.002 | 0.754 ±0.002 |
| This work | **0.071** ±0.001 | **0.693** ±0.004 | **0.682** ±0.005 | **0.361** ±0.098 | **0.694** ±0.007 | **0.782** ±0.003 | **0.374** ±0.115 | **0.843** ±0.003 | **0.836** ±0.004 | **0.811** ±0.004 |

### 3.2. Performance evaluation

In Table 5, we compare the assessment performance of the proposed method with other top-of-the-line ones which can be categorized into either single-aspect assessment, or multi-aspect and multi-granular assessment. The first two rows show the single-aspect assessment methods, viz. the deep feature method [13] and an SSL-based method [25]. The former uses a feature derived from the acoustic model instead of GOP to obtain an utterance-level holistic score, while the latter adopts the HuBERT Large features to predict either the fluency or the prosody scores. Although both of them can gain some improvements over GOP, they limit themselves to the single-aspect assessment, which may be insufficient to evaluate the speaking proficiency of L2 learners. On the other hand, the recently proposed methods (viz. LSTM [14], GOPT [14] and 3M [15]) can generalize well to the multi-aspect and multi-granular assessment, especially when integrating the segmental and suprasegmental features together [15]. Conversely, instead of using parallel modeling paradigms, the last two rows are hierarchical methods (viz. HiPAMA [26] and our method). In comparison to our method, HiPAMA has a fall-off in the overall performance, this may attribute to the lack of using suprasegmental features and the sub-word level modeling. On the other hand, our method leads to better assessments across all granularities and surpass the other methods compared in this paper. The results also demonstrate that the modeling of the hierarchical architecture is more amenable to the rendering intricate traits of speech when compared with 3M [15]. Furthermore, due to the synergistic effects of integrating the sub-word- and word-level modeling, our method can markedly improve the performance of various word-level assessments.

Table 2. *Ablations studies on word-level scores.*

| Setting | Word Score (PCC) ↑ | | |
|---|---|---|---|
| | Accuracy | Stress | Total |
| This work | **0.682** ±0.005 | 0.361 ±0.098 | **0.694** ±0.007 |
| - DS-Conv | 0.642 ±0.011 | 0.327 ±0.167 | 0.655 ±0.013 |
| - Relative Position | 0.637 ±0.009 | **0.433** ±0.032 | 0.651 ±0.008 |
| - Sup-phoneme | 0.633 ±0.006 | 0.319 ±0.203 | 0.649 ±0.004 |

### 3.3. Ablation studies: sub-word and word-level modeling

On the ground of the distinct improvements of the proposed method on the word-level scores shown in Table 1, in the last set of experiments, we conduct ablation studies on sub-word- and word-level modeling to examine the underlying factors. It can be seen from Table 2 that DS-Conv contributes the most to the overall scores, which is consistent with the results of a word-level MDD task [18]. Therefore, in this work, we reaffirm that adopting the convolution operation is crucial in CAPT, which can not only capture the local context cues but also the sub-word level mispronunciation characteristics. Furthermore, the introduced sup-phone embeddings that complements the canonical phone embeddings can also help promote the performance of word-level assessments.

## 4. Conclusion

In this paper, we have put forward a novel context-aware hierarchical modeling method for multi-aspect and multi-granular APA. A series of experiments conducted on the speechocean762 benchmark dataset have demonstrated the feasibility of this promising method in comparison to several top-of-the-line methods. In the future, we plan to explore more discriminative features in our model and also envisage enhancing the model robustness for practical applications [27].

---

[1] https://www.openslr.org/resources/11/librispeech-lm-corpus.tgz